\documentclass[%
 prl,
 amsmath,amssymb,
twocolumn,
]{revtex4-2}

\usepackage{graphicx}% Include figure files 
\usepackage{dcolumn}% Align table columns on decimal point
\usepackage{bm}% bold math
\usepackage{color}
\usepackage{comment}
\usepackage[utf8]{inputenc}
\usepackage[T1]{fontenc}
\usepackage{mathptmx}

\usepackage{todonotes}

\begin{document}

\title[Intermittent dynamic bursting in vertically vibrated liquid
drops]{Intermittent dynamic bursting in vertically vibrated  liquid drops} 
% Force line breaks with \\

\author{Andrey Pototsky}
 \affiliation{Department of Mathematics, Swinburne University of
   Technology, Hawthorn, Victoria, 3122, Australia}%Lines break
                                %automatically or can be forced with
                                %\\ 
\email{apototskyy@swin.edu.au}
\author{Ivan S.~Maksymov}%
%\todo{\Ivan{Just added my patronymic initial.}}
\affiliation{Optical Sciences Centre, Swinburne University of
  Technology, Hawthorn, Victoria, 3122, Australia\looseness=-1%\\This
                                %line break forced with
                                %\textbackslash\textbackslash 
}
\author{Sergey A.~Suslov}
\affiliation{Department of Mathematics, Swinburne University of
  Technology, Hawthorn, Victoria, 3122, Australia} 
\author{Justin Leontini}
\affiliation{Department of Mechanical and Product Design Engineering,
  Swinburne University of Technology, Hawthorn, Victoria, 3122,
  Australia} 

\date{\today}

\begin{abstract}
  A previously unreported regime of type III intermittency is
    observed in a vertically vibrated milliliter-sized liquid drop
    submerged in a more viscous and less dense immiscible fluid layer supported by a hydrophobic solid plate.
As the vibration amplitude is gradually increased, subharmonic Faraday waves are
  excited at the upper surface of the drop.
  We find a narrow window of vibration amplitudes slightly above the
  Faraday threshold, where the drop exhibits an irregular sequence of large
  amplitude bursting events alternating with intervals of low amplitude activity.
The triggering physical mechanism is linked to the competition between surface Faraday waves and the shape deformation mode of the drop.
\end{abstract}

\maketitle
Intermittency is an intricate dynamical regime that is observed in
many driven dissipative systems.
It is broadly associated with irregular alternation
between intervals of a low-disorder quiescence phase and a high-disorder bursting phase.
Theoretically described in the context of
dynamical systems in \cite{Pomeau80}, intermittency presents
  itself as an initial instability of the low-disorder phase (usually
  a periodic regime). Instabilities associated with local bifurcations of 
periodic solutions---saddle-node, torus or period-doubling---give
rise to type I, II and III intermittency, respectively. Different
types of intermittency have been experimentally observed, for
  example, in Rayleigh-B\'enard convection \cite{Dubois83, Ciliberto88},
Taylor-Couette flow \cite{Swinney83}, brain activity \cite{Grant94},
electro-chemical reactions \cite{Okamoto98, Basset87},  chemical reaction systems with two slow parametric excitations  \cite{Zhou20} and in
thermoacoustic oscillations of flame dynamics \cite{Basu18,Kabiraj12}. 
\begin{figure}
\includegraphics[width=0.99\hsize]{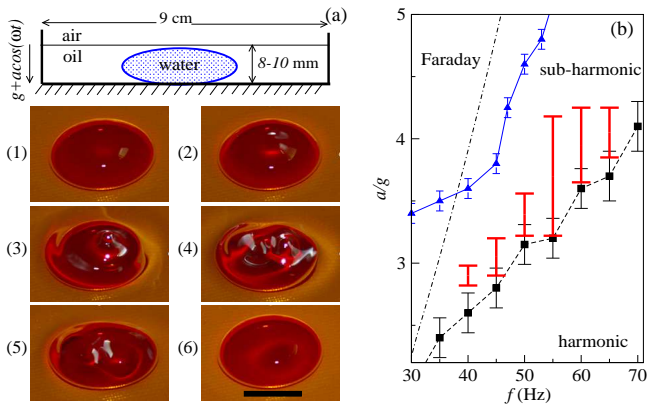}  
\caption{\label{F1} (a) Schematic diagram of the experimental setup: a
  water drop submerged in an oil film and supported by a hydrophobic
  silicone wafer. 
In the co-moving frame of the vertically vibrated
  plate the acceleration is $(g+a\cos(\omega t))$. (a.1--a.6) A single
  bursting event of a $1.5$\,mL water drop in $9$\,mm olive oil
  film, vibrated at $f=50$\,Hz with the amplitude $a=3.3g$. The time
  interval between two subsequent snapshots is $1$\,s and the scale
  bar in (6) is $10$\,mm. (b) Phase diagram of dynamical regimes. Below the dashed line (
  black squares) the drop  exhibits small amplitude harmonic
  oscillations at the forcing frequency without any visible changes in
  its shape. The squares mark the onset of the subharmonic Faraday waves at
  the top of the drop. The thick red error bars indicate the regions, where
  intermittent bursts were found. The triangles represent the
  threshold for the onset of Faraday waves on the surface of the
  $9$\,mm oil film in the $90$\,mm circular Petri dish {\it without} the
  drop. The dashed-dotted line is a theoretical threshold for
  the Faraday waves on the oil layer {\it without} the drop obtained
  according to \cite{Kumar96} for an infinite horizontal layer of the
  $9$\,mm-thick oil layer over a no-slip substrate. 
}
\end{figure}

An example of a system with a rich bifurcation scenario is a
vertically vibrated liquid drop surrounded by a layer of a more
viscous immiscible fluid. At a certain critical vibration amplitude, the
drop undergoes a period-doubling bifurcation that corresponds
to the onset of subharmonic Faraday waves on its surface
\cite{Pucci11, Pucci13, Ebata15}. If the drop is floating in a
more dense fluid, the period-doubling bifurcation gives rise to a new
stable configuration in which the drop elongates to form a worm-like
structure \cite{Pucci11, Pucci13}. However, when the drop is
immersed in a fluid of similar density (i.e.~an almost neutrally
buoyant drop), the variety of dynamic states born as the result of the 
primary period-doubling bifurcation becomes remarkably rich. 
This includes walking drops driven by the surface Faraday waves, rotating
drops and stationary drops with irregular polygonal shapes
\cite{Ebata15}. However, intermittent dynamic regimes have
  been thus far not reported in such systems.
    
Here, we report for the first time on experimental
observations of a type III intermittency in a vertically vibrated
milliliter-sized drop submerged in a more viscous and less dense
immiscible carrier fluid. The experiments are performed
  using a water or Soybean sauce drop immersed in vegetable or olive oil. Typical
  oil dynamic viscosity, density and oil-air surface tension are 
$\mu=84$\,mPa$\,$s, $\rho=916$\,kg$/$m$^3$ and
$\sigma=0.035$\,N$/$m, respectively.  The water viscosity and density
are $\mu_w=1$\,mPa$\,$s and $\rho_w=1000$\,kg$/$m$^3$, respectively.
 Soybean sauce is heavier and more viscous than water with density
 $1080$\,kg$/$m$^3$ and viscosity between $1.6$ and $3.6$ \,mPa$\,$s
 \cite{soy}. The 
newly-found intermittent regime, when the drop transits between a quiescent phase of
small-amplitude oscillations and violent
bursts of large amplitude waves on its upper surface, is very delicate. It exists only when the drop
height is less than, but almost equal to, the thickness of the oil
layer, i.e.~the drop is fully submerged with its highest point almost
touching the oil surface. In the experiments with a 30--60\,ml oil
  bath we found that changing the overall volume of the oil by as
  little as 1\,mL could trigger or destroy the bursting regime. If
  the drop is covered by the oil layer that is thicker than the
  capillary length $\sqrt{\sigma/\rho g}\approx 2$\,mm of oil, the
  bursting regime is no longer found.

  Such high sensitivity of the bursting regime to the thickness of oil film
  cap above the drop is not surprising: it is known since the pioneering
  work of Benjamin Franklin \cite{Franklin1774} that oil has a
  ``calming'' effect on water ripples. Recent studies demonstrate 
  that even a monolayer of oil ($\sim 10$\,nm) dramatically suppresses
  capillary waves on water due to the Gibbs-Marangoni effect
  \cite{Behroozi07}.

Experiments were conducted with water or Soybean Sauce
  drops with a horizontal diameter 10--15\,mm supported by a
hydrophobic Teflon substrate placed on the bottom of a 90\,mm glass
Petri dish. The drop was submerged in an 5--10\,mm deep
oil layer, which was sufficient to just cover the drop, see Fig.~\ref{F1}(a).

The dish was mounted on a $6.5$'' $45$\,W RMS audio speaker (Sony,
Japan) powered by a $30$W stereo amplifier (Yamaha TSS-15, China) of a
sinusoidal signal produced by  a digital tone
  generator. The dish was vertically vibrated at frequencies in the range
40--70\,Hz. The vibration amplitudes were sufficiently large to excite
subharmonic Faraday waves on the surface of the drop but not strong
enough to excite Faraday waves on the free surface of the oil layer.

The acceleration amplitude $a$ of the Petri dish was measured with the
accuracy of $\pm 0.08g$ using an ADXL\,326 accelerometer
(Analog Devices, USA) connected to a digital
oscilloscope (Tektronix TDS 210, USA). The motion of the
 drop was recorded using an in-house photodetector
measuring the intensity of light reflected by the drop
surface.  In case of the water drop, a small amount of red food dye
  was added to increase contrast with the yellow oil and a DC LED lamp was used as a light source to avoid
flickering in the range of the investigated driving frequencies. 

Our main result is the discovery of a narrow window of vibration
amplitudes, slightly above the period-doubling bifurcation,
where the drop exhibits a random sequence of large-amplitude
bursts alternating with long intervals of
  small-amplitude oscillations.
During a
bursting event (Fig.~\ref{F1}(a.1--a.6)) a localised wave develops
on the upper surface of the drop. Its wavelength is comparable
with the horizontal drop size. The wave crest develops within the first
1--2\,s in the centre of the drop and reaches a magnitude of surface
deflection of up to 5\,mm. After about 1--2\,s the initial circular
wave is replaced by an irregular polygonal surface wave that
eventually dies out and the drop re-enters another interval of
small-amplitude harmonic oscillations at the forcing frequency with
no visible change in shape of the drop 
%\footnote{See the slow motion
%  video of a 1.5\,mL water drop vibrated at $50$\,Hz with amplitude
%  $a=3.3g$ in the supplemental material}. 
We note that the  
dynamics of the surface waves during the burst is not unique
and generally depends on the drop volume and the driving frequency.

Fig.~\ref{F1}(b) shows the results of scanning the $(f,a)$ plane by
gradually increasing or decreasing the vibration acceleration
amplitude $a$ by the smallest possible amount of $0.08g$ that
corresponds to the measurement error at fixed frequency for a
$1.5$\,mL water drop.  

Below the
dashed line connecting the black squares, the drop behaves as a
forced oscillator: it slightly wobbles at the forcing frequency
without any visible change in shape. As $a$ is increased 
above the dashed line, subharmonic Faraday waves are
excited on the upper surface of the drop. Significantly, we found that
the dynamics of the drop close to the dashed line is
  multi-stable with the harmonic and subharmonic regimes
coexisting. Thus, when $a$ is gradually increased, the subharmonic
waves set in at the upper end of the error bars
  associated with the black squares. Alternatively, when $a$ is
gradually decreased, harmonic oscillations are reinstated
at the lower end of the  error bars associated with
  the black squares.

We find that the intermittent bursts exist in the frequency range
  between 40 and 65\,Hz at amplitudes larger than those corresponding
  to a subharmonic transition shown by the thick red vertical
  bars in the figure.
  Note that an overlap between the intermittency and
  multi-stability amplitude ranges exists only at the bottom of an
  intermittency range. Therefore, the intermittency is not driven by the
  interaction of periodic and subharmonic oscillation states.

As the amplitude is increased further, the intermittency is destroyed
and the drop enters a new dynamic regime characterized by persistent
subharmonic surface waves and a horizontal drop elongation
%\footnote{See the slow motion video of a 1\,mL water drop vibrated at
%  $60$\,Hz with amplitude $a=4.7g$ in the supplemental
%  material}. 
These 
post-intermittency states are similar to those found in water drops
swimming in silicone oil \cite{Ebata15} and in a floating liquid lens
\cite{Pucci11, Pucci13}.

To validate our experimental results we compare the onset of Faraday
waves in a $9$\,mm-thick oil layer {\it without} the drop with the exact
theoretical threshold for an infinitely extended horizontal layer on a
no-slip substrate \cite{Kumar96}. The corresponding experimental
(theoretical) threshold is shown by the triangles (dashed-dotted line)
in Fig.\ref{F1}(b). 

\begin{figure}
\includegraphics[width=0.99\hsize]{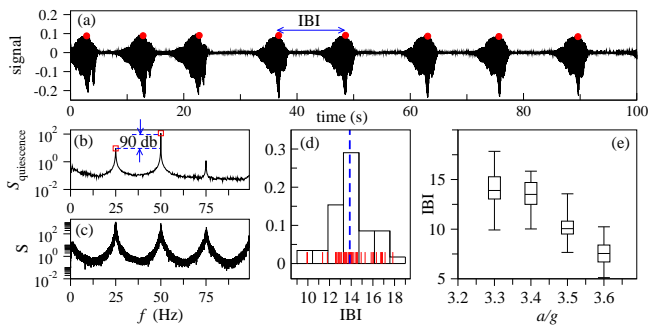} 
\caption{\label{F2} Intermittency in a $1.5$\,mL water drop in the
  $9$\,mm-thick olive oil layer vibrated at $50$\,Hz with the
  amplitude $a=3.3g$. (a) Signal recorded by the photodetector over
  $100$\,s showing a sequence of bursts (spikes) alternating
  with quiescent phases of small amplitude
    oscillations. The inter-burst intervals (IBI) are
    determined by detecting the peaks shown by circles during the
    bursts. (b) Power spectrum of the quiescent phases. The
  dominant harmonic peak at $50$\,Hz is $90$\,dB stronger 
  than the sub-harmonic peak at $25$\,Hz. (c) Power spectrum of the
  entire signal recorded over $10$\,min. (d) Distribution of the
  inter-burst intervals from (a). (e) Dependence of the IBI distribution on the
  vibration amplitude $a$ at $50$\,Hz. Box plots
  summarise the IBI distributions for  $a=3.3g$, $3.4g$, $3.5g$ and
  $3.6g$.}
\end{figure}
Fig.~\ref{F2}(a) shows a representative signal recorded by
the photodetector.
The occurrence of a burst is identified by
locating the largest peak in the amplitude (shown by the filled
circles). The burst duration is well-defined and is approximately
  $5$\,s while the inter-burst intervals (IBI) are
highly irregular
with the average of
$13.9$\,s and the inter-quartile range of $15.3-13=2.3$\,s, see
Fig.~\ref{F2}(d).  We emphasise that the statistics of the
duration of the quiescent phases are similar to that of 
the inter-burst intervals because the duration of bursts
is approximately 5\,s.

We recorded the signal over $10$ minutes and identified the quiescent
phases (low disorder intervals) as times when the oscillation
amplitude is below $10\%$ of the maximum spike
amplitude. Figure~\ref{F2}(b) shows the typical power spectrum
of a quiescent phase, which is dominated by the
sharp harmonic peak at $50$\,Hz. The second strongest subharmonic
peak at $25$ Hz is $90$\,dB weaker than the harmonic one. On the
contrary, the power spectrum of the entire signal is dominated by a
broad-band subharmonic peak at $25$\,Hz and its higher harmonics at
$n\times 25$\,Hz ($n=2,3,\dots$), see Fig.~\ref{F2}(c). This result
is indicative of the type III intermittency, namely:
the bursts occur as a result of the period-doubling
bifurcation from the quiescent phase (low disorder state of harmonic
oscillations).

The existence of intermittency in the 40--70\,Hz frequency
range was confirmed in experiments with 0.8--1.5\,mL water and
0.5--1\,mL Soybean Sauce drops 
%\footnote{See the real-time video in
%  the supplemental material showing intermittency in a 0.5\,mL Soybean
%  Sauce drop in 5mm oil layer vibrated at $65$ Hz with amplitude
%  $a=4g$}. 
This frequency range is dictated by the requirement that 
  the Faraday wave length has to be comparable with the diameter of
  the drop. Larger drops tend to elongate under the action of Faraday
  waves so that a new dynamic regime is reached with no intermittency
  found.

Notably, when the vibration amplitude is increased further, the
intermittency is destroyed, but the drop does not necessarily enter a
  chaotic regime. For example, for a $1.5$\,mL water drop vibrated at
  $50$\,Hz we first found intermittent bursts at $a=3.3g$. We
  subsequently increased the vibration amplitude to $a=3.4g$, $3.5g$ and
  $3.6g$ recording the signal for $10$ minutes and extracting the IB
  intervals. The data summary presented in Fig.~\ref{F2}(e) shows a
  gradual decrease of the duration of the quiescent phase with
  increasing amplitude. When the vibration amplitude was increased
  beyond $a=3.6g$, the intermittency was destroyed and the drop 
  elongated by the Faraday wave entered a new highly regular state
  characterised by subharmonic oscillations similar to the
  states reported in \cite{Pucci11, Pucci13}.

To elucidate the physical mechanism that triggers the
intermittent bursting regime found in this work, we closely inspect the snapshots of the
bursting phase from Fig.\ref{F1} and slow motion videos presented
in Supplementary Materials. In the quiescence phase the drop appears as
perfectly circular, when viewed from the top. At the beginning of a
bursting phase, Faraday waves developing on the upper surface of the
drop break the axial symmetry and excite the shape deformation
mode. Using the analogy with spherical harmonics $Y_{l}^{n}$, the
shape deformation mode can be associated with the oscillation between
a prolate and oblate spheroid, i.e.~$l=2$ and $n=0$.  As the bursting
phase develops further, the deformation of the drop increases and
competes with Faraday waves to eventually suppress them and push the
drop back into the quiescent phase.  

Note that such a mode competition mechanism has been observed to trigger
chaos in earlier experiments with nonlinear Faraday waves excited by
vibration in a circular fluid layer \cite{Cilberto84}. More recently,
in the experiments with floating liquid drops, the interplay between
Faraday waves and deformability of the drop boundaries was shown to
lead to their mutual adaptation, when a stable elongated shape of the
drop is reached \cite{Pucci11, Pucci13}. Here we observed that a
similar mode competition leads to a new dynamical regime of repeating
bursting events in a heavier drop surrounded by a more viscous and
less dense fluid layer.

We propose a simple phenomenological model of intermittent bursting,
given by system Eqs.\,(\ref{eq2a},\ref{eq2b}) that takes into account the
coupled dynamics of two oscillation modes: the Faraday surface wave with
amplitude $F$ and a shape deformation mode with amplitude $v$.
\begin{eqnarray}
  \ddot{F}+(c_1 +d v^2)\dot{F}+(w_0^2+A\cos{(\Omega t)})F+F^3
  &=&b_1\cos{(\Omega t)}\,,\label{eq2a}\\ 
  \ddot{v} +c_2\dot{v}+(\omega_0^2+F)v+v^3&=&0\,,\label{eq2b}
\end{eqnarray}
The Faraday mode $F$ follows a damped and driven
Mathieu equation Eq.~(\ref{eq2a}) with cubic nonlinearity
that describes the three-wave
interaction process known to give rise to triangular
broad-band power spectra in nonlinear Faraday waves
\cite{Punzmann09, Shats12, Maksymov19}. 
The frequency of the forcing is $\Omega=2$
in the units of the eigenfrequency of the subharmonic Faraday mode
$w_0=1$. The shape deformation mode $v$, on the other hand, has a much
smaller natural frequency $\omega_0 \ll 1$ than the Faraday mode and
is not directly excited by forcing, as it can be seen in the
slow-motion videos in the supplement, which clearly show that the drop
almost ``freezes'' between bursting events. 

To couple the two modes, we assume that the shape deformation mode $v$
is linearly excited by the Faraday mode $F$ and that the latter is
suppressed by $v$ via a quadratic damping nonlinearity, given by the
term $d v^2\dot{F}$. Other model parameters include the intrinsic
damping coefficients $c_1$ and $c_2$, the parametric forcing
amplitude $A$ and the direct forcing amplitude $b$.

\begin{figure}
\includegraphics[width=0.95\hsize]{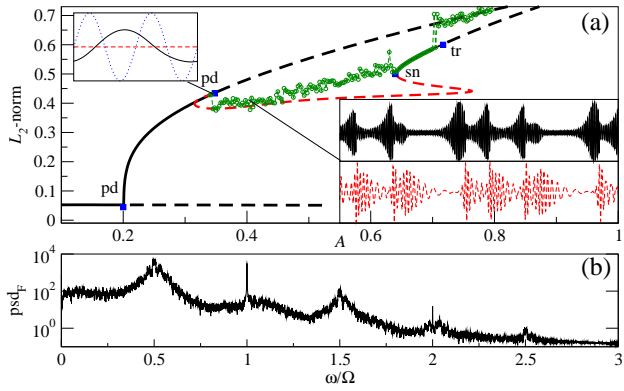} 
\caption{\label{F4} (a) Bifurcation diagram of system (\ref{eq2a}) and
   (\ref{eq2b})  with a varied forcing amplitude $A$ for
   $c_1=0.1$, $d=1$, $b_1=0.1$, $c_2=0.05$ and
   $\omega_0^2=0.01$. 
%\SAS{Can $\omega_0$ be set to 0 and omitted from the model?} 
   The thick solid (dashed) lines correspond to stable (unstable) solution
   branches. The green circles are obtained from the direct numerical
   integration of (\ref{eq2a}) and (\ref{eq2b}). The labels pd, sn and tr denote the period-doubling, sadle-node and torus
     bifurcations, respectively. The insets in (a) show the solutions
   at the period-doubling bifurcation point and at $A=0.4$, as
   indicated by arrows. The solid, dashed and dotted lines correspond
   to the Faraday mode, the shape deformation mode and forcing,
   respectively. (b) Power spectrum (psd) of the Faraday mode $F$ at
   $A=0.4$. 
}
\end{figure}

We study periodic solutions of the system (\ref{eq2a},\ref{eq2b}) 
using a numerical continuation method \cite{AUTO}. The
bifurcation diagram obtained by varying the forcing amplitude $A$ is
shown in Fig.~\ref{F4}(a). The first period-doubling (pd) bifurcation
at $A=0.2$ corresponds to the onset of the Faraday waves at the upper
surface of the drop without the excitation of the shape deformation mode, as
shown in the inset of Fig.~\ref{F4}(a). This point corresponds to the
dashed line in Fig.~\ref{F1}(b). Slightly above the primary
period-doubling bifurcation, Faraday waves undergo a secondary
subcritical period-doubling bifurcation at $A=0.35$. At this point, the
drop enters the regime of irregular pulsations
represented by the sequence of amplitude spikes in both the
Faraday and volumetric modes (see the inset in Fig.~\ref{F4}(a)). This
non-periodic regime can only be found by numerical integration of
(\ref{eq2a}) and (\ref{eq2b}). The time-averaged $L_2$-norm of
the solution is shown by the green circles
 in Fig.~\ref{F4}(a). The power spectrum of the Faraday mode
contains one sharp peak at the driving frequency $\omega=\Omega=2$ and
a broadband background as shown for $A=0.4$ in Fig.~\ref{F4}(b). 

\begin{figure}
\includegraphics[width=0.95\hsize]{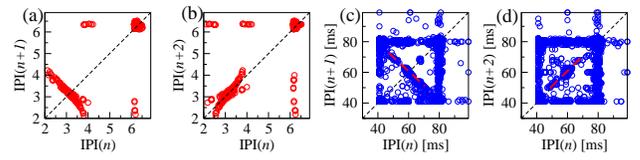} 
\caption{\label{F5} (a, b) The first and the second return maps of the
  inter-peak intervals in the Faraday mode  in the model
  (\ref{eq2a},\ref{eq2b}) at $A=0.4$. (c, d) Return 
    maps of the inter-peak intervals IPI$(n)$ with duration larger
    than 40 ms, for the experimental data from Fig.~\ref{F2}(a). The
    dashed line shows the bisector.}
\end{figure}

To identify the type of the intermittent bursting regime, we extract
all peaks from the time series of the Faraday mode amplitude
$F(t)$ at $A=0.4$. The first and the second return maps of the
inter-peak intervals IPI$(n)$ are shown in Fig.~\ref{F5}(a,b). We
focus on the quiescent phase that is characterised by the inter-peak
intervals close to $\pi$. The first return maps of the quiescent phase
appear to cross the bisector line while the second return maps are
parallel to it.

This feature is a signature of the type III intermittency
\cite{Schuster06}. We find a similar feature in the return maps of the
inter-peak intervals with the duration larger 
than $40$\,ms extracted from the experimental data series shown in
Fig.~\ref{F5}(c,d) 
\footnote{The return maps of the inter-peak intervals in the
  experimental data with duration less than 30\,ms are too noisy and
  cannot be used to reveal any hidden structure.}.

To conclude, we have observed experimentally intermittent dynamic
  bursting in a vertically vibrated liquid drop surrounded by a less
  dense and more viscous liquid layer and supported by a hydrophobic
  solid plate. 
  The physical mechanism triggering such a regime is identified as
  a competition between the Faraday mode and the shape deformation
  mode that is responsible for  the horizontal elongation of the drop.  
  The universal nature of the phenomenon is confirmed by detecting
  intermittency in organic (Soybean Sauce) and inorganic (water)
  liquid drops. The occurrence of the intermittent bursting regime is
  found to be highly sensitive to the variations of the oil layer
  thickness above the drop that diminishes during bursting events due
  to the action of large-amplitude Faraday waves mixing of the
  microscopic quantities of the drop liquid with oil.

AP would like to thank Krzysztof Stachowicz for technical support and
Evgeny Mogilevsky for stimulating discussions. IM
  acknowledges support by the Australian Research Council through
  Future Fellowship (Grant No. FT180100343).

%

%\bibliography{bibfile}
\end{document}